# A Semantic Grid-based E-Learning Framework (SELF)

Zaheer Abbas[+], Muhammad Umer[+], Mohammed Odeh[*], Richard McClatchey[*], Arshad Ali[+], Farooq Ahmad[+]

[+]*National University of Sciences & Technology (NUST)*
*NUST Institute of Information Technology*
*Rawalpindi 46000, Pakistan*
*{zaheer,m.umer,drarshad,farooq.ahmad}@niit.edu.pk*

[*]*CCCS Research Centre, University of the West of England (UWE)*
*Frenchay, Bristol BS16 1QY, United Kingdom*
*{mohammed.odeh,richard.mcclatchey}@uwe.ac.uk*

**Abstract**

*E-learning can be loosely defined as a wide set of applications and processes, which uses available electronic media (and tools) to deliver vocational education and training. With its increasing recognition as an ubiquitous mode of instruction and interaction in the academic as well as corporate world, the need for a scaleable and realistic model is becoming important. In this paper we introduce SELF; a Semantic grid-based E-Learning Framework. SELF aims to identify the key-enablers in a practical grid-based E-learning environment and to minimize technological reworking by proposing a well-defined interaction plan among currently available tools and technologies. We define a dichotomy with E-learning specific application layers on top and semantic grid-based support layers underneath. We also map the latest open and freeware technologies with various components in SELF.*

## 1. Introduction

E-learning can be defined as a 'wide set of applications and processes, which use available electronic media (and tools) to deliver vocational education and training' [1]. More recently, an alternative definition of E-learning by Alexander J. Romiszowski defines E-learning as a 'solitary, individual activity, or a collaborative group activity (where) both synchronous (real-time) and asynchronous (flexi-time) communication modes may be employed' [2]. From web-based learning to innovations such as online conferences, E-learning has progressed a long way. World-wide interest in E-learning can be seen from the estimates of the E-learning market growing to 50 billion USD by 2010 from a current market of around 3 billion USD [3]. Historically, the Internet and the World Wide Web gave birth to the concepts of E-learning and collaborative knowledge sharing across the globe, but due to largely unplanned and unanticipated growth, are now falling short of earlier promises. Lack of machine readable content coupled with information overload has put strains into the traditional knowledge delivery model of WWW. The situation is especially serious in the E-learning domain where the success and usefulness directly correlates with the effectiveness of knowledge delivery in a dynamic setting. A large number of research efforts are hence focusing on a planned infrastructure development for E-learning [4,5,6]. Our work is a major step in the same direction but offers a more integrated approach.

The hypotheses behind our work are: (i) a scaleable E-learning infrastructure requires an inherent support for heterogeneity and (ii) integration of available technology using a well defined plan is the optimal path for achieving a practical E-learning infrastructure. In this paper we introduce SELF; a Semantic-grid based E-Learning Framework. Rather than proposing a strategy of development from scratch and ending-up with yet another monolithic architecture with low integration capabilities, we suggest an integrated approach that involves minimal re-work of existing systems. We firstly identify the key enablers for a realistic E-learning infrastructure and map these to available technologies to establish a well-defined framework for the integration of these technologies so that the goals of effective E-learning can be achieved.

The rest of this paper is organized as follows: Section 2 covers the key enablers in E-learning infrastructure and a discussion on the delivery mechanism, section 3 provides an overview of SELF and its various layers; section 4 reviews technologies for integration in SELF, Section 5 is a brief review of related work while conclusions and references are presented in sections 6 and 7.

## 2. The Semantic Grid in E-learning

The basic architecture behind the World Wide Web (WWW) is not capable of providing a seamless and artificially-intelligent framework required for a large scale and effective E-learning implementation. Hence, several research efforts worldwide are focusing on resolving this issue.

The concept of grid computing support for E-learning has long received criticism from various quarters. Critics hold the view that the modern day WWW architectures, tools and technology support nearly every feature required by E-learning and thus incorporating the grid is largely unnecessary for E-learning support. We are studying this assertion taking into consideration the evolved form of Computer Supported Collaborative Learning (CSCL) around the world supported by grid computing. Table 1 presents a round-up of a basic set of enablers for an effective E-learning program that we extracted from a study of major CSCL projects around the world, specifically SOUL [7], AccessGrid[5], APPLE[8], OntoEdu[26] and ConferenceXP[9]. Table 1 also gives a subjective ranking (as per the authors' point of view) of WWW support for these enablers.

**Table 1: Key enablers for effective E-Learning infrastructure**

| Effective E-learning Enablers | WWW Support |
|---|---|
| Seamless sharing of large pool of resources (information, storage, customized software/hardware and computational power) | Average |
| Support for a dynamic and continuously evolving set of participants | Average |
| Support for Service Oriented Architecture | High |
| Support for dynamic content and resource management | Low |
| Intelligent indexing/match-making for resources and contents | Average |
| Standards for security and trust | Average |
| Collaborative tools for groupware management | High |
| Knowledge Management | Average |
| On-Demand QoS | Low |

The analysis presented in Table 1 is founded on the fact that the support for the above mentioned enablers is not catered for in current WWW architectures. Even though some technologies have allowed the WWW to accommodate current workloads, but no technological support can make up for architectural shortcomings.

While it may be obvious that WWW can support E-learning on a limited scale, any extensive implementation will require significant architectural changes. This is precisely the point where the grid fits neatly into this paradigm. It is intuitive to map the basic E-learning enablers onto the features inherent in a grid infrastructure. The enablers in table 1 are not specific for E-Learning only. They are also suitable for general grid applications.

The support for dynamism in terms of resources, content and participants may be considered as a core grid-based architectural feature to support effective E-learning strategies. A closer look into any E-learning based infrastructure would identify the highly dynamic pattern of its key ingredients i.e. resources (both content and computation) and participants. The management of such dynamically changing environments, being a key task of grid computing, needs to be further extended seamlessly to E-learning environments.

Since both the grid and the WWW hold certain strengths as far as E-learning support is concerned, we propose to create a synergy by incorporating the so-called 'semantic grid' in E-learning. The semantic grid merges the semantic web with grid computing. Also, incorporating semantic grid in E-learning will provide the best seamless support available through a merger of the best of both paradigms. In the next section we propose a semantic grid based architecture for E-learning and a discussion of its major strengths.

## 3. A Semantic-Grid based E-learning Framework (SELF)

The proposed semantic grid-based framework is presented in Figure 1. A service oriented semantic grid middleware lies at the core of SELF. A Service Oriented Architecture (SOA) will enable a loosely coupled inter-relationship between Collaborative Partners (CPs) and provide a higher abstraction in the form of open interfaces. We propose a layered SELF stack at each Virtual Organization (VO) with two major segments corresponding to both E-learning and semantic grid applications/services. Such a layered approach gives a better understanding of interaction between the various components. This layered SOA approach allows us to decouple the independent components of SELF. We present below a brief functional description of each of these layers.

### 3.1. E-learning applications layer

The top layer in the SELF VO stack will carry various end-user applications such as group and courseware managers, search facilities, scheduling and tracking software etc. Of course all of these applications will be dependent on and controlled by the specific requirements of respective end users. Possible examples of such applications could fall in the domains of E-Teaching, E-Training, E-Workshop and E-Conference.

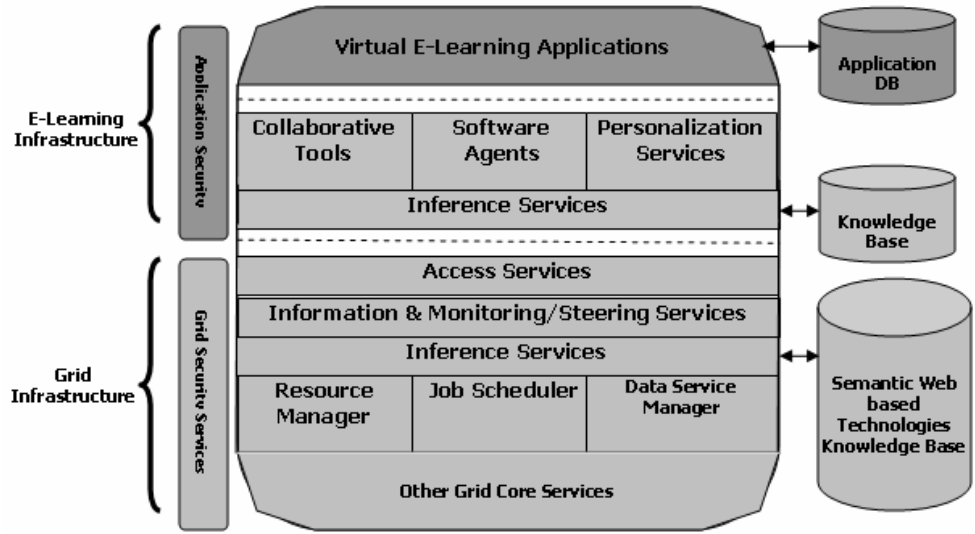

**Figure 1: Semantic Grid based E-Learning Framework (SELF)**

### 3.2. E-learning applications layer

The top layer in the SELF VO stack will carry various end-user applications such as group and courseware managers, search facilities, scheduling and tracking software etc. Of course all of these applications will be dependent on and controlled by the specific requirements of respective end users. Possible examples of such applications could fall in the domains of E-Teaching, E-Training, E-Workshop and E-Conference.

### 3.3. E-learning services layer

The development of applications will be facilitated by a set of generic application-level services such as collaboration tools, agents and personalization managers. A recent application of collaboration tools and services can be seen in CoAKTing [6], which provides services such as the status of collaborative partners, discussion minutes, meeting status, things to do list, project status etc. The benefit of decoupling applications and services into separate layers is twofold. First, it will minimize re-working and increased maintainability of applications by the result of high cohesiveness and loose coupling. Second, it will also ensure compliance to some standardization criteria during application development.

Personalization Services (PS) may impart an important role by personalizing the individual centric information. That is, if someone is interested in lecture materials of some special domain, then the PS executing as a backend process will both keep track of such information based on the content usage and reduce the latency involved in information retrieval. Also, the PS can be deployed at the site level so that each individual's required information can be kept up-to-date. It could also change the traditional learning processes from strong push delivery, lack of personalization and the linear/static learning processes to efficient, distributed, student-oriented, personalized, and non-linear/dynamic learning processes [29]. Readers interested in a detailed comparison of the characteristics of traditional learning process vis-à-vis E-learning process are referred to [29]. The PS will be based on specific policies and indexing approaches determined by the interest of users (either defined explicitly by the users or inferred through usage patterns).

A common theme among E-learning applications in a collaborative environment is to provide intelligent search, matching and inference support. Our basic hypothesis in this regard is that a small set of generic inference services can cater for a large number of applications. A typology of the proposed inference services is shown in figure 2.

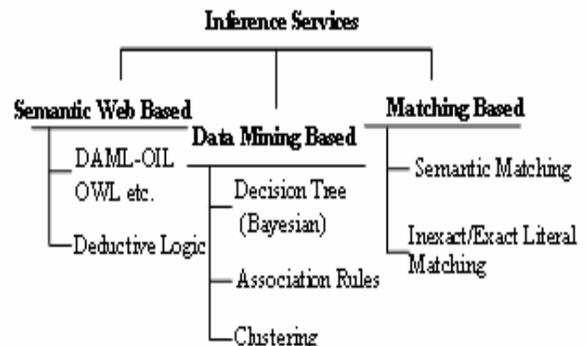

**Figure 2: A typology of inferences services**

Resource and content matching becomes a major issue since the use of a description language still does not standardize services and content description to the extent where a direct string-based matching can be applied. We believe that the required matching fits nicely in to the problem of semantic data matching from the information theory paradigm. Also known as in-exact data matching, semantic data matching as the name suggests is matching on the basis of semantics or meanings of data rather than its character or literal formation. Probably the most successful application of semantic data matching is in the web search engines where terms and documents are co-clustered on the basis of semantics.

Latent Semantic Analysis (LSA) is a powerful semantic matching technique [30, 31]. It is used to extract semantic similarity between pieces of textual information using statistical techniques. Similarly, there also exist other techniques, which involve graphical classification of the concepts and then semantic matching based on set theory and AI concepts [32]. A significant body of work could be found in the literature where such semantic or in-exact matching techniques are proposed or applied to the grid domain [11, 12]. Although it is very challenging to integrate all the inference services together in one framework because of the interoperability issues, the advantages are that it will further strengthen the SELF for information retrieval.

To ensure appropriate support for the inference services each VO must maintain a knowledge base (e.g. for educational applications a repository of lectures, videos, tutorials, experiment designs and results and proof of experiments etc.). This may be done by deploying an artifact management system [10], which will maintain the documentation, processes, researcher or tutor profiles for future reuse. Since different ethnic groups from heterogeneous locations will be sharing their logical resources in a collaborative E-Learning environment, there must be some comparable standardized form of lectures, tutorials, videos etc. This could enable Collaborative Partners to be compliant within a shared environment. Some standards like the Educational Modeling Language (EML) from the Open University of Netherlands, IMS Enterprise Services for managed learning environments [10] already exist.

The following list may be considered as a minimal set of parameters required for an "exact" information search approximately equivalent to the IEEE Learning Object Metadata (LOM) model. These parameters could best be defined using ontologies from the semantic web domain.

a. *Domain:* the major category of related material, whether it belongs to medical sciences, computer sciences, astronomy or any other domain.

b. *Type:* the type of document e.g. ppt, pdf, doc, avi, wav etc.

c. *Author: the* author(s) who generated a resource.

d. *Size/Capacity:* the total size of a resource. If it is a document then this is measured in bytes or if it is some computational resource then its computation power in Hz.

e. *Location:* the address/location of the resource.

f. *Description:* the short description of a resource.

g. *Constraints:* the basic requirements like security restrictions or specific tools to execute a specific request. This list can be extended based on domain-specific requirements of the users.

### 3.4. The Semantic Grid and SELF

The strongest and most innovative component of SELF is its core component, the semantic grid. When several VOs are participating, the problems of heterogeneity and low standardization limit the applicability of a conventional grid.

The potential benefits of the semantic grid approach over conventional grids can be numerous. In a conventional grid infrastructure, VOs are required to agree upon a framework for resource description so that the grid services can share, locate and apply security measures [11, 12]. In a dynamically changing and evolving environment, such a requirement constrains the scalability of this framework. This is exactly where we anticipate that the semantic grid can help in overcoming the resource description problem using model-theoretic solutions from the semantic web domain. The semantic grid envisions a well-structured integration of semantic web and classical grid technologies.

In a recent work, Goble & De Roure provide three possible options to deal with dynamism and heterogeneity in a semantic grid environment [13]. These include knowledge management techniques, semantic grid services and Multi-Agent Systems (MAS). Among these options, MAS is probably the most innovative where intelligent software agents can negotiate and translate descriptions within the collaborating environment. Agents can also play a key role in extracting important information from required resources. Agents could also enable certain advance features such as masking-off the heterogeneity of structured information and support for autonomous asynchronous operations without affecting the normal workflow of other processes.

### 3.5. Security infrastructure in SELF

Security issues such as the authorized access to shared resources, the conservation of intellectual property rights, the confidentiality of contents, the authentication of individuals, and the auditing of resource usage play an important role in the smooth and controlled working of any distributed system. In a grid environment, it is difficult to manage the workflow of the complete system in a secure manner. Despite the abovementioned security requirements, several VOs which can become partners and form virtual markets in a single grid environment, might have different security policies and mechanisms to provide secure and controlled access to resources. Therefore a common mechanism which can mask the heterogeneity of security policies is needed either in the form of wrappers (as in GRASP [33]) or common policy description languages (as in XACML [34]). Furthermore, we believe that in SELF, we need a security infrastructure to support both the underlying core of grid (Grid Security Infrastructure – GSI) and its collaborating applications layer using security add-on components.

**Table 2: Proposed tools & technologies for SELF**

| SELF Component | Proposed Tool/Technology |
|---|---|
| Collaborative Tools | CoAKTing [6] |
| Personalization Services | CHANDLER [17] |
| Inference Services | OILEd, JESS, Description Logic |
| Software Agents | SAGE [18], JADE [19], Aglets [20] |
| Grid Components and Services | JClarens [16,21], GT3[22] |
| Ontologies | Protégé etc. |
| Semantic Web Techniques | OWL[23], DAML-OIL[24] |
| Security | Assorted Security Add-ons |

## 4. Enabling Technologies for SELF

As the popularity of the grid grows, a large number of open tools and technologies are being developed world-wide. Table 2 specifies some open tools and technologies that we plan to use in the implementation phase. Not all of the tools or technologies are required for specific components, rather these are the multiple implementation options available.

Figure 3 presents an abstract deployment model of SELF, showing the components needed at each site. The following abstract example explains the workflow of SELF with the help of a suitable scenario. Assume that all participants from different sites have deployed all the layered components needed for SELF. Suppose one user from site A intends to go through lectures/contents of some specific subject. His/Her knowledge query by using SELF layers will be given to software agents as well as personalization services to keep track of user interests.

Software agents by communicating with collaborative partner agents using inference services will resolve query for demanding resources based on the parameters outlined in section 3.2. Software agents and inference services will have to access the grid resource descriptions through access services, information monitoring and management services.

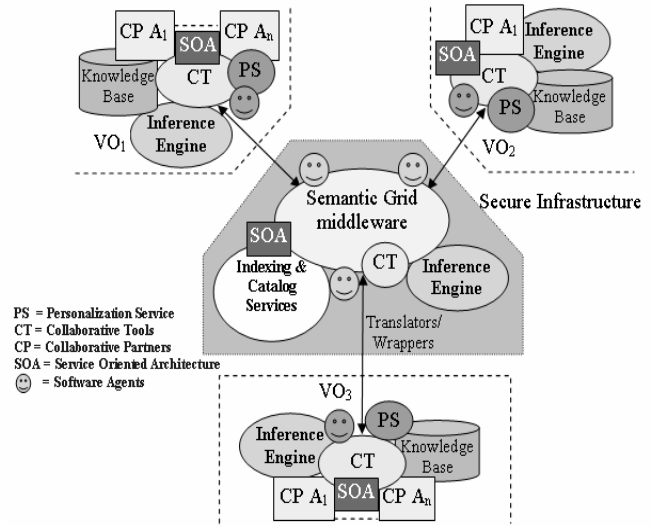

**Figure 3: An abstract deployment model for SELF**

This access to agents will also be provided based on the inference services specific to knowledge base of grid middleware, which will keep track of the resources descriptions and their usage status. It seems very trivial but it is very difficult to integrate all the technologies mentioned in figure 1 and table 2. Important and critical aspect of the SELF is its adaptive learning, updating the knowledge bases for future requests and collaborative tools for meetings and participants' information etc. In the proposed implementation scheme, we have purposely limited ourselves to open source technologies only. The biggest implementation challenge is to integrate various technologies in an efficient and scalable manner. Precedents of such integration could be found in works such as the integration of web services with multi agent systems [14, 15], the integration of collaboration tools in grid environment [6] and the incorporation of security features in JClarens (VOMS; Virtual Organization Membership Service) [16]. The SELF implementation results will enable the suitability of SELF for CSCL to be evaluated.

## 5. Related work

With increasing intellectual and commercial collaborations across the globe, the E-learning domain is a rapidly developing field as demonstrated by the number of technologies that we have referred to in this paper. It is therefore important to locate the current work with respect to other research and it is in this light that we review some of the recent works which have informed and motivated the current reserach.

### 5.1. OntoEdu: Ontology-based Education Grid System for e-Learning

OntoEdu, a project of the University of Peking, China, is the most recent work wherein an ontology based grid is proposed for educational applications [26]. Using educational technologies at its crux, the OntoEdu architecture realized concept reusability with ontology, device and user adaptability with ubiquitous computing and automatic composition. Although the OntoEdu architecture is quite innovative and extensive in nature, its primary focus is oriented towards adaptability (personalization). Designers have not referred to the incorporation of some equally important areas such as special collaboration tools or services and intelligent search and matching agents.

### 5.2. CoAKTing: Collaborative Advanced Knowledge Technologies in the Grid

CoAKTing provides a motivating example of the incorporation of collaborating technologies on top of a grid structure [6]. Geared towards academic and intellectual collaborations, CoAKTing is a set of collaborating tools that enables enhanced process tracking and navigation of resources before, after, and during meetings in progress [6, 28]. These tools work through a shared ontology and could be integrated in an existing collaborative environments (such as the Access Grid [5]). Each of the CoAKTinG tools can be thought of as extracting structure from the collaboration process [28]. The CoAKTing project has introduced tools such as BuddySpace for presence awareness, Compendium for keeping track of a bundle of ideas, issues and conceptual interrelationships involved in projects, I-X Process Panels and Meeting Replay. Interested readers are referred to [6] and [28] for more details on working and features of these tools.

The set of CoAKTing tools can be useful at collaboration services layer of SELF. We agree, with [28] where they summarize that 'the CoAKTinG tools (if not specific tools, the concepts underlying them) can be transposed into the Learning Grid'.

### 5.3. APPLE: A Novel P2P based e-Learning Environment

The APPLE (A novel P2P based e-Learning Environment) project emphasizes the importance of grid and P2P infrastructures for e-Learning applications instead of a static web [8]. This work proposes the use of the grid for group-centric and P2P for individual-centric information retrieval. The designers of APPLE used WSRF.NET to develop and deploy a virtual classroom service. They integrated a P2P platform with the grid to exploit extensive resource potential from the grid. Despite being an extensive framework, a major limitation of APPLE seems to be its dependency on a proprietary Microsoft technology (WSRF.NET). Moreover, intelligent semantic matching structures, personalization and collaboration technologies have not been explicitly addressed in the original APPLE proposal.

In a larger sense, the use of P2P (as in APPLE) or the grid (as in OntoEdu, CoAKTing and SELF) have similar final objectives — the pooling and coordinated use of large sets of distributed resources. These technologies work with the same approach to solving the problems but target different communities, resources and applications. In an important paper, Foster and Iamnitchi state that the complementary nature of the strengths and weaknesses of the two approaches suggests that the interests of the two communities (grid and P2P) are likely to grow closer over time [27]. In the same spirit, the designers of APPLE incorporate the strengths of both technologies by adopting a hybrid architecture. The SELF philosophy further enhances this approach by introducing the semantic grid based underlying middleware with reasoning support for easy service discovery and request submission, software agents for intelligent negotiation and collaborating tools for the purpose of collaborative activities like meetings, things to do list etc.

### 5.4. Other Related Works

An exciting work in collaborative learning is the Access Grid project of Argonne National Labs [5]. Currently deployed at 150 institutions around the world, the Access Grid is a multicast videoconference technology that enables its users to conduct real-time virtual conferences and maintain a wholesome online groupware.

Boldyreff et. al. have explored the concept of shared artifacts over the grid [10]. All the resources, such as documentation including architectural details, design documents, test cases, process definitions and details, researcher or partner profiles are considered to be the artifacts which can be shared for future reuse over the grid. Boldyreff et. al. also made an effective analogy between collaborative software development and collaborative learning by highlighting the significance of shared artifacts over the grid.

Some recent works have also been reported in the domain of the integration of semantic web technologies either in the form of deploying translators [11] or using ontology based matchmakers [12]. The induction of semantics in the grid will further improve the collaborative efforts in different domains. Large scale projects including DILIGENT and BRICKS are underway for the integration of digital libraries for collaborative heterogeneous knowledge sharing within grid environments [25].

## 6. Conclusions

An effective, end-to-end and practical E-learning environment cannot be realized from a loose integration of available technologies or by starting the development from scratch. The former approach could lead to an unrealistic and non-scalable infrastructure, while the latter strategy might end up with wasteful rework. A rather efficient approach requires, (i) an understanding of key enablers behind the target E-learning infrastructure, (ii) a comparative analysis of available tools and technology on the basis of customizability, applicability and cost, (iii) a mapping of key E-learning enablers onto the available technology, and (iv) a detailed architecture that specifies the interaction among the technological solutions at various levels.

In this paper, we have attempted to evolve an end-to-end E-learning infrastructure from the integration of available technologies, specifically the semantic web, the grid, collaborative and personalization tools, and knowledge management techniques. We understand that it will be very challenging to integrate all the components of SELF into a single framework because of interoperability issues. The implementation of SELF will follow a bottom up approach to better understand and mask the heterogeneity issues of multiple tools and technologies. A review of recent research and development work in this domain suggests the need for the development of such a large number of tools that may be used at various levels in the proposed SELF architecture. Finally, the outcomes of this research at this initial stage may be regarded as a step forward to disseminate ideas on a proposed semantic and grid-based architecture for the effective understanding, integration, and deployment of E-learning applications based on a proposed framework for E-learning.

## Acknowledgements

The authors wish to acknowledge the support of their home institutions and the funding bodies that have supported this work. Special thanks is extended to colleagues that have stimulated ideas in the field of E-Learning frameworks.

## 7. References

[1] Eklund, J., Kay M. & Lynch H.M.. "E-Learning: Emerging Issues and Key Trends: A Discussion Paper", Australian National Training Authority (ANTA), 2003, available at http://flexiblelearning.net.au/research/2003/elearning250903final.pdf

[2] Romiszowski, A, "How's the E-learning Baby? Factors Leading to Success or Failure of an Educational Technology Innovation", Educational Technology, Vol. 44, No. 1, January-February 2004, pp. 5–27

[3] Levis, K., "The Business of E-learning: A Revolution in Training and Education Markets", 2002, Report summary available at http:// www. hrmguide. net/usa/hrd/e-learning_survey. htm

[4] Wilson S., Blinco k. & Rehak D., "An e-Learning Framework: A Summary", A Paper prepared on behalf of DEST (Australia), JISC-CETIS (UK), and Industry Canada, July 2004. available at http://www.jisc.ac.uk/uploaded_documents/Altilab04-ELF.pdf

[5] Home page of Access Grid project, http://www.accessgrid.org/

[6] Shum, S.B., "CoAKTinG: Collaborative Advanced Knowledge Technologies in the Grid", Proc. Second Workshop on Advanced Collaborative Environments, Eleventh IEEE Int. Symposium on High Performance Distributed Computing (HPDC-11), July 24-26, 2002, Edinburgh, Scotland.

[7] Space Online Universal Learning (SOUL) homepage, http://www.soul.hkuspace.org/home/eng/index.html

[8] Hai Jin et al., "APPLE: A Novel P2P based E-Learning Environment", 6th International Workshop on Distributed Computing (IWDC 2004), Indian Statistical Institute, Kolkata, India, 27-30 December, 2004

[9] Introducing ConferenceXP, Microsoft ConferenceXP homepage, http://www.conferencexp.net/community/default.aspx

[10] Boldyreff, Cornelia, et. al., "Towards Collaborative Learning via Shared Artefacts over the Grid", in proceedings of 1st LEGE-WG International Workshop on Educational Models for GRID Based Services, 16 September 2002, Lausanne, Switzerland.

[11] Brooke J., et al., "Semantic Matching of Grid Resource Descriptions", in proceedings of 2nd European Across-Grids Conference, 2004.

[12] Harth, Andreas et. al., "A Semantic Matchmaker Service on the Grid". In proceedings of 13th International World Wide Web Conference (WWW2004), New York, May 17-22 2004, Pages 326-327.

[13] Goble, Carole & De Roure, David, "The Semantic Grid: Myth Busting and Bridge Building," in proceedings of 16th European Conference on Artificial Intelligence (ECAI-2004), Valencia, Spain, 2004

[14] Farooq A. et al., "Autonomous Distributed Service System: Basic Concepts and Evaluation", in proceedings of Grid and Cooperative Computing, Second International Workshop (GCC 2003), Shanghai, China, December 7-10, 2003, pp. 432-439.

[15] Suguri, Hiroki et. al., "AgentWeb Gateway--Enabling Service Discovery and Communication among Software Agents and Web Services. in proceedings of. Third Joint Agent Workshops and Symposium (JAWS2004), pp. 212-218, October 2004, Karuizawa, Japan.

[16] Arshad Ali et. al, "JClarens: A Java Based Interactive Physics Analysis Environment for Data Intensive Applications", in proceedings of the International Conference of Web Services (ICWS'04), San Diego, USA, 2004.

[17] Open Source Application Foundation (OSAF), "What's Compelling About Chandler: A Current Perspective", available at http://www.osafoundation.org/Chandler_Compelling_Vision.htm

[18] Ghafoor A. et. al., "SAGE: Next Generation Multi-Agent System", in proceedings of PDPTA, Navada, USA, 2004, pp. 139-145, Vol. 1.

[19] Multi-Agent Systems: JADE resources: WWW Link: http://jade.tilab.com/

[20] Mobile Agents – Aglets resources: WWW Link: http://aglets.sourceforge.net/

[21] JClarens, Assorted Resources: WWW Link: http://clarens.sourceforge.net/jclarens/list.html

[22] Globus Tool Kit 3, Assorted Resources: WWW Link: http://www-unix.globus.org/toolkit/

[23] Web Ontology Language (OWL) Overview available, http://www.w3.org/TR/owl-features/

[24] DAML-OIL, Assorted Resources: WWW Link: http://www.daml.org/2001/03/daml+oil-index.html last accessed in December 2004.

[25] Frommholz, Ingo et. al., "Supporting Information Access in Next Generation Digital Library Architectures", in proceedings of the Sixth Thematic Workshop of the EU Network of Excellence DELOS, Cagliari, Italy, June 2004 pp. 49-60.

[26] Guang-zuo CUI et al., "OntoEdu: Ontology based Education Grid System for e-learning", GCCCE2004 International conference, Hong Kong. 2004

[27] Foster, Ian and Iamnitchi, Adriana, "On Death, Taxes, and the Convergence of Peer-to-Peer and Grid Computing", 2nd International Workshop on Peer-to-Peer Systems (IPTPS'03), February 2003, Berkeley, CA.

[28] Bachler, Michelle et al., "Collaboration in the Semantic Grid: a Basis for e-Learning", in proceedings of the 7th International Conference on Intelligent Tutoring Systems Workshop (ITS'2004), Maceio, Brazil, August 30th, 2004.

[29] Stojanovic L., Staab S. and Studer R., "eLearning based on the Semantic Web", In proceedings of World Conference on the WWW and Internet (WebNet2001), Orlando, Florida, USA, October 23-27, 2001.

[30] Deerwester S. et. al., "Indexing by latent semantic analysis", Journal of the Society for Information Science, 41(6), pp. 391-407, 1990.

[31] Berry M.W., Dumais S.T. and O'Brien G.W., "Using Linear Algebra for Intelligent Information retrieval", SIAM review, 37(4) pp. 573-595, 1995.

[32] Deborah L. et al., "Towards Explaining Semantic Matching", Technical Report # DIT-04-019, Proceedings of the Int. Workshop on Description Logics, February 2004.

[33] Chadwick D. et al., "Multilayer Privilege Management for Dynamic Collaborative Scientific Communities", UK Workshop on Grid Security Practice, Oxford, July 2004.

[34] "eXtensible Access Control Markup Language (XACML) Version 1.0 3", OASIS Standard, 18 February 2003, available at http://www.oasis-open.org/committees/xacml/repository/